\documentclass[10pt,a4paper,twocolumn]{article}

\usepackage{graphicx}
\usepackage[explicit]{titlesec}
\usepackage{authblk}
\usepackage{multirow}
\usepackage{etoolbox}
\usepackage{soul}
\usepackage{lipsum}
\usepackage[labelfont=bf,labelsep=endash]{caption}
\usepackage{balance}
\usepackage{tabularx}
\usepackage{xcolor}
\usepackage[hidelinks]{hyperref}
\usepackage[hang]{footmisc}
\usepackage[normalem]{ulem}
\usepackage[top=1in, bottom=1in, left=0.69in, right=0.69in]{geometry}
\usepackage[T1]{fontenc}
\usepackage{newtxmath,newtxtext}
\usepackage{mathtools}
\usepackage[none]{hyphenat}

\titleformat{\section}{\center\normalfont\bfseries}{\thesection.}{1em}{\MakeUppercase{#1}}
\titlespacing*{\section}{0pt}{12pt}{9pt}

\titleformat{\subsection}{\normalfont\bfseries}{\thesubsection}{1em}{#1}
\titlespacing*{\subsection}{0pt}{12pt}{9pt}

\titleformat{\subsubsection}{\normalfont\itshape}{\thesubsubsection}{1em}{#1}
\titlespacing*{\subsubsection}{0pt}{12pt}{9pt}

\newcommand{\ITUurl}[1]{\textcolor{blue}{\urlstyle{same}\url{#1}}}

\def\starttable{\vspace{6pt}\begin{table}[ht]\center}
\def\startfigure{\vspace{6pt}\begin{figure}[ht]\center}

\makeatletter
\def\tagform@#1{\maketag@@@{\ignorespaces#1\unskip\@@italiccorr}}
\makeatother

\title{\large{\textbf{\uppercase{Experimental demonstration of a real time wideband OFDM generation in SUB-THZ}}}}

\author{\normalsize{Eray Güven, Nesrine Benchoubane and Güneş Karabulut Kurt \\ Poly-Grames Research Center, Department of Electrical Engineering, Polytechnique Montréal, QC, Canada}}

\date{}

\begin{document}

\maketitle

\begin{abstract}
\textit{Sub-terahertz (Sub-THz) wireless communications and their potential applications continue to attract significant attention and foster debate on the usage of their unused frequency bands to relieve existing spectrum congestion. However, for these next-generation networks, experimental research is limited by the lack of flexible, real-time testbeds. This study presents a real-time, multi–radio-frequency (RF)-channel, cascaded software-defined radio (SDR)-based Orthogonal Frequency-Division Multiplexing (OFDM) transmission platform achieving an aggregate sampling rate of $2 \times 3.84$ GSPS and approximately $1.1$ GHz of instantaneous bandwidth, targeting sub-THz and THz SDR testbeds. The digital transmitter architecture, including the OFDM signal processing chain and instrumentation workflow, is described in detail. A comparative case study between a conventional sub-6 GHz implementation and a 180 GHz configuration is conducted, evaluating phase noise, spectral occupancy, received average and peak power. The direct impact of sub-THz bandpass filtering, necessitated by harmonic-mixer-based upconversion, is also experimentally analyzed. Measurement results show conversion and filtering losses of up to $24.1$ dB, while the system exhibits stationary phase noise levels on the order of $-60$ dBc/Hz, demonstrating the feasibility and limitations of real-time wideband OFDM transmission at $180$ GHz. Beyond immediate current capabilities, the platform builds a foundation for the scalable integration of multiple transmitters and receivers, which is essential for the implementation, conformance, and testing of emerging sub-THz communication systems.}
\end{abstract}






\section{Introduction} 
\label{sec:intro}

Subterahertz (Sub-THz) communication has emerged as a promising solution to spectrum congestion and the demand for ultra-high-data-rate wireless links. Since the introduction of Long-Term Evolution (LTE), orthogonal frequency-division multiplexing (OFDM) has become a well-established multicarrier waveform for wideband communications and has consequently been adopted in numerous THz and sub-THz experimental testbeds reported in the literature \cite{Yamaguchi2024, 9798100, Liu2024_HighSpeed022THz}.

In contrast to hardware-defined, fixed-function sub-THz testbeds, software-defined radio (SDR) platforms provide enhanced flexibility, reconfigurability, and sustainability, enabling sub-THz communication research to transition from application-specific implementations toward programmable and reusable experimental platforms \cite{9798100, 7820226, 10580085}. One recent study has highlighted both the opportunities and challenges of SDR-based THz systems \cite{10622496}, noting in particular that many existing testbeds lack real-time baseband data generation, thereby limiting efficient utilization of the sub-THz and THz spectrum.

To address scalability and bandwidth limitations, multi-radio-frequency (RF)-channel cascaded architectures based on Radio Frequency System-on-Chip (RFSoC) platforms have recently been introduced for mmWave and sub-THz applications \cite{10580085, csahin2023millimeter}. In this work, we investigate a real-time multi-RF-channel OFDM transmission system at 180 GHz based on an FPGA-customized RFSoC platform. Furthermore, while harmonic mixers are widely employed for RF-based THz upconversion in such testbeds, the resulting multi-harmonic interference necessitates bandpass filtering, whose direct impact on spectral distortion and power attenuation is systematically analyzed in this study. 

The main contributions of this study are summarized as follows:

\begin{itemize}

\item A real-time, multi–RF-channel, cascaded SDR-based OFDM transmission platform achieving an aggregate sampling rate of $2 \times 3.84$ GSPS and over 1 GHz of instantaneous bandwidth is presented as a sub-6 GHz implementation.

\item A hybrid digital signal processing (DSP)–hardware architecture is developed, combining flexible digital signal processing with high-speed RF front-end implementation to enable wideband real-time operation.

\item An improved sub-6 GHz implementation using a 180 GHz SDR platform is demonstrated and compared, with observed spectral discrepancies analyzed and explained.

\item The impact of sub-THz bandpass filter (BPF) frequency selectivity is experimentally investigated, quantifying its effects on power attenuation and in-band spectral clipping of the dual band OFDM channels.
\end{itemize}

The remainder of this paper is organized as follows. Section 2 and Section 3 presents the RFSoC-based transmitter architecture and OFDM DSP chain respectively. Section 4 introduces the experimental sub-THz testbed, while Section 5 evaluates the system under sub-6 GHz and 180 GHz operation, followed by concluding remarks in Section 6.

\begin{figure*}[t]
    \centering
    \includegraphics[width=\linewidth]{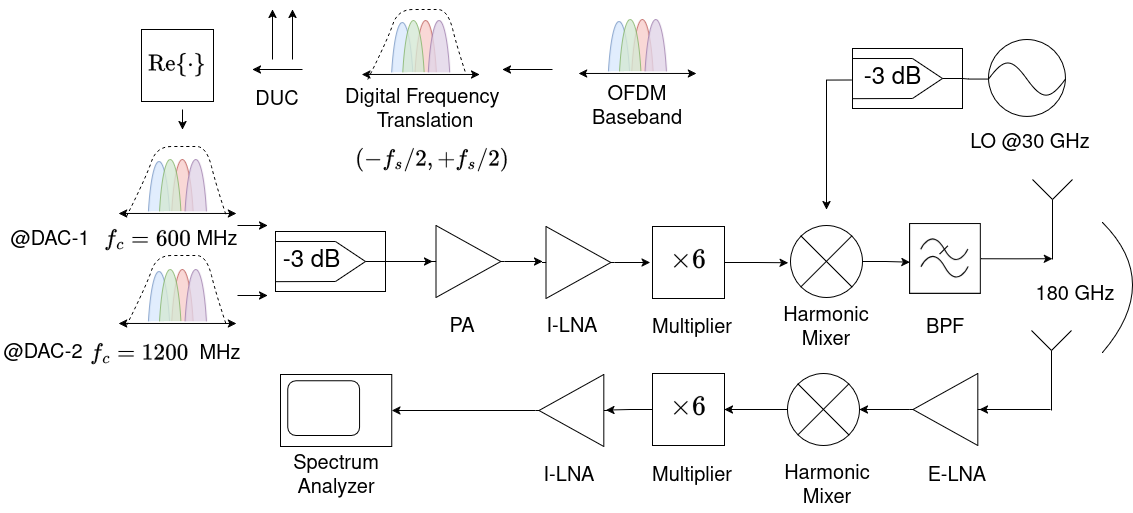}
    \caption{Block diagram of the experimental setup.}
    \label{fig:block_diagram}
\end{figure*}
\begin{figure*}
    \centering
    \includegraphics[width=0.9\linewidth]{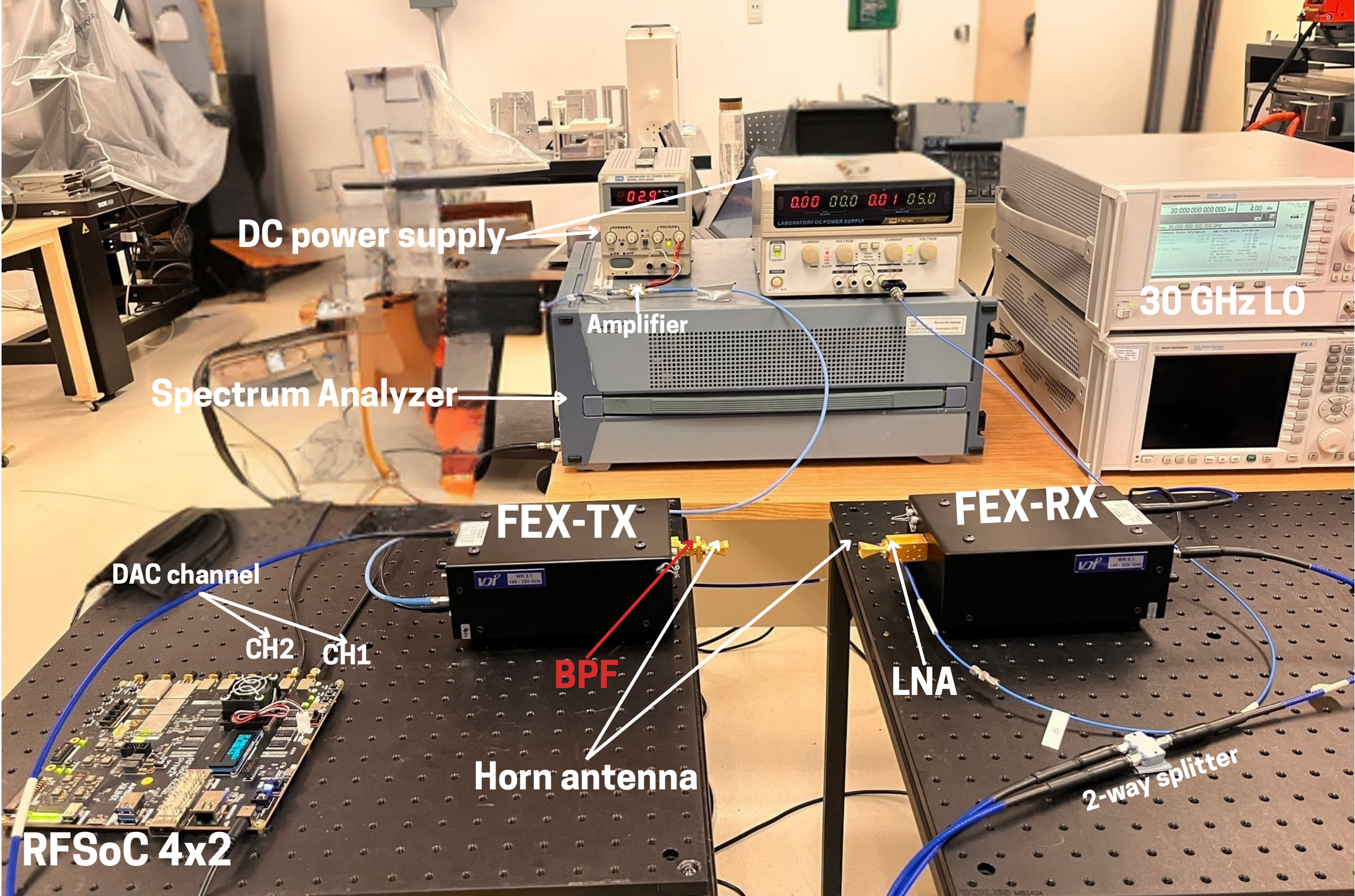}
    \caption{Overview of the experimental setup. }
    \label{fig:overview}
\end{figure*}

\newpage\section{Digital Design}
The digital-to-analog converter (DAC) tiles are equipped with integrated baluns, enabling single-ended RF outputs. The RF Data Converter IP controls two DAC channels using independent AXI-Stream interfaces. Each DAC operates at a sampling rate of $3.84$ Giga-Samples per Second (GSPS), where the maximum achievable rate is $7$ GSPS. A fabric clock of $192$ MHz is used to generate a DAC tile phase-locked loop (PLL) reference frequency of 340 MHz independently for each tile. The AXI4-Lite control interface between processing system (PS) of RFSoC and FPGA registers operates at $100$ MHz bus clock. In RF Data Converter (RFDC), $10 \times$ interpolation stage is enabled in the digital up-conversion (DUC) chain, where four samples are forwarded per AXI-Stream clock cycle at $192$ MHz.

The programmable logic design, including the overlay configuration, is implemented using the Vivado Design Suite 2020.2, while RF Data Converter IP customization is performed in MATLAB R2020a and Vitis HLS where carrier aggregation for FDM is coordinated. A commercial off-the-shelf (COTS) host computer is used for programming and overlay management outside runtime. PYNQ is a Python API framework for PL access, enabling RFDC configuration. During runtime, register based back communication from PS to PL via AXI4-Lite is used for operating frequency hopping, and high speed data transfer does not exist. This connection is done via USB 3.0 Micro-B. 

\section{Digital Signal Processing (DSP)}

Each OFDM symbol contains $N = 64$ subcarriers, comprising $48$ data subcarriers, $4$ pilot subcarriers, and $12$ zero-padded subcarriers. The subcarrier spacing is thus $\frac{20 ~ \text{MHz}}{64} = 312.5 ~ \text{kHz}.$ A $16$-point cyclic prefix (CP) is appended to each OFDM symbol. The complete OFDM frame structure is provided in the accompanying open-source repository \cite{rfsoc_ofdm}. After digital-to-analog conversion, the DAC output exhibits Nyquist-zone spectral replication, resulting in 24 total spectral images of the 20 MHz OFDM waveform across the $\pm f_s$ analog frequency span. With a $3.84$ GSPS DAC sampling rate, the first Nyquist zone extends to $1.92$ GHz, which is sufficient to accommodate the combined bandwidth of the two DAC channels.
Consequently, $ 64 \times 24 = 1536$ subcarriers are present at the DAC output, occupying an effective bandwidth of $480$ MHz centered at a configurable RF frequency. This center frequency is set using the digital direct synthesis (DDS) mixer integrated within the RFDC and configured via the XRFDc library on PYNQ \cite{Xilinx_RFSoC-PYNQ}.

\section{Experimental setup}

The full block diagram illustrating the different stages is shown in Fig.~\ref{fig:block_diagram}.
The experiment, as shown in Fig~\ref{fig:overview}, is conducted in an indoor laboratory environment using the (SDR)–THz testbed at the Poly-Grames Research Center \cite{polygrames2025}. An AMD Xilinx Radio Frequency System-on-Chip (RFSoC) 4×2 device, hereafter referred to as RFSoC-TX, is used as the transmitter.

For upconversion from intermediate frequency (IF) to sub-THz and downconversion from sub-THz to IF, Virginia Diodes, Inc. (VDI) frequency extenders (FEXs) are employed. FEX-TX operates as a transceiver, while FEX-RX is configured as a receiver. Each FEX incorporates a frequency multiplication factor of $M = 6$, implemented as a cascaded frequency doubler and tripler. With a $30$ GHz local oscillator (LO) input, the RFSoC IF signal is upconverted to a $180$ GHz carrier frequency. Each FEX is equipped with identical circularly polarized rectangular horn antennas, exhibiting 3 dB beamwidths of $8.9^\circ$ and $10.28^\circ$ in the vertical and horizontal planes, respectively.

In addition, a full-band bandpass filter (BPF) covering 140–220 GHz is inserted at the FEX-TX front end to investigate the impact of out-of-band interference suppression in the sub-THz link. The filter exhibits a –20 dB cutoff frequency range of 135–230 GHz with an average insertion loss of 0.75 dB \cite{vadiodes}. Each IF channel is amplified using a Mini-Circuits ZX60-V63 power amplifier, providing 19 dB gain in linear operation region. At the receiver, the sub-THz front end of FEX-RX employs a low-noise amplifier (LNA) with 20 dB gain.

The received spectrum is measured using a Rohde and Schwarz FSIQ40 spectrum analyzer. A Keysight E8267D pulse signal generator (PSG) provides the $30$ GHz LO for both FEX-TX and FEX-RX via a two-way splitter with a $10$–$40$~GHz bandpass response. The IF amplifiers are powered by two independent DC power supplies operating at $\le 5$ V.

\section{Experimental results}

\begin{table*}[htbp]
\centering
\renewcommand{\arraystretch}{2} 
\caption{Comparative output power and phase noise for CH1 and CH2 across the different measurements.}
\label{tab:power_phasenoise}
\begin{tabular}{llccc}
\hline
\textbf{Measurements} & \textbf{Channel} & \textbf{Output Power (dBm)} & \textbf{Phase Noise (dBc/Hz)} & \textbf{Output}  \\ \hline
\multirow{2}{*}{Sub-6 GHz} 
                      & CH1 (600 MHz) & -40.0 & -65 & \multirow{2}{*}{\includegraphics[width=0.1\textwidth]{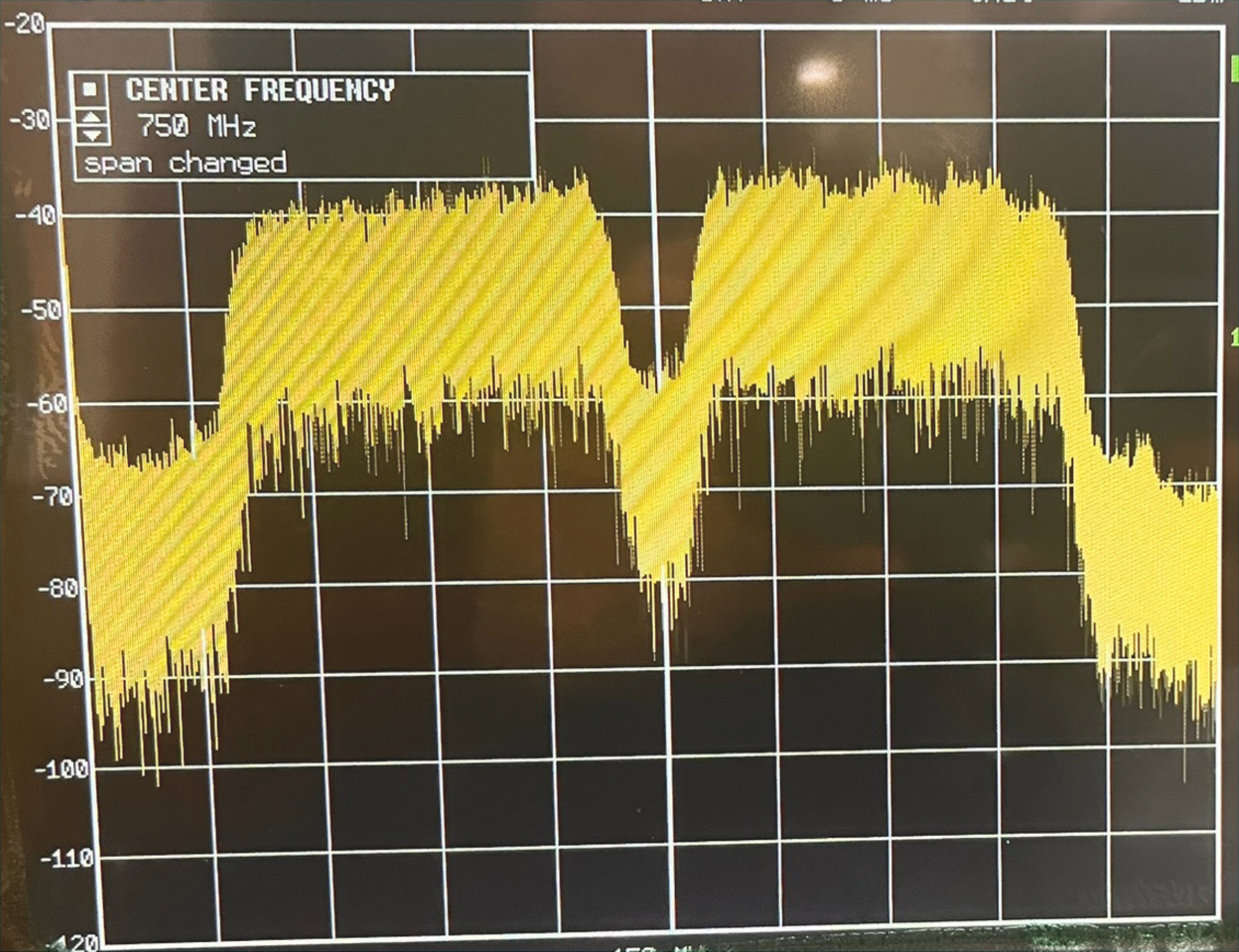}}\\ & CH2 (1200 MHz) & -40.0 & -63 &   \\ \hline \noalign{\vskip 5pt} 
\multirow{2}{*}{$180$ GHz with BPF} 
                      & CH1 (600 MHz) & -64.1 & -63 & \multirow{2}{*}{\includegraphics[width=0.1\textwidth]{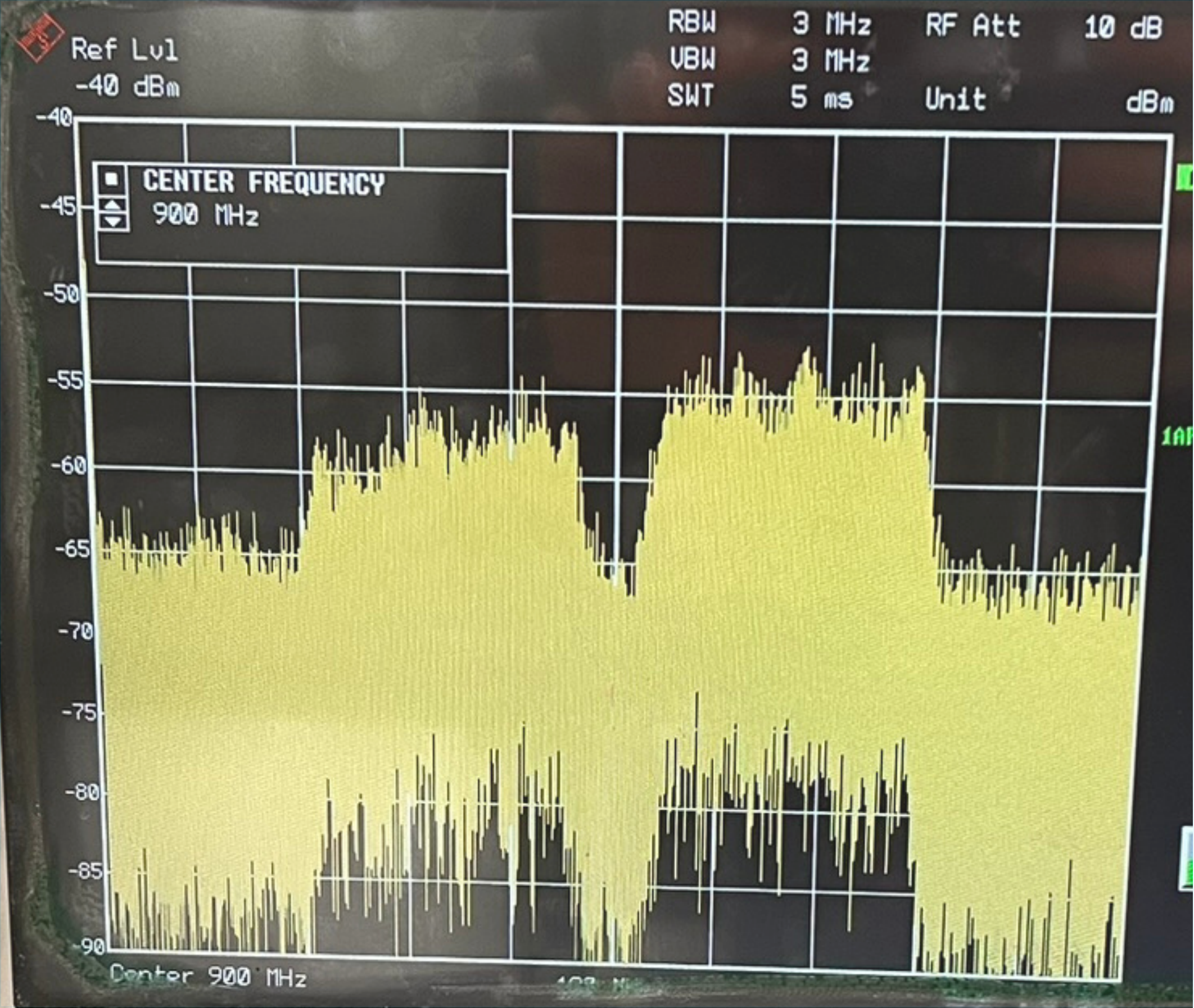}} \\  & CH2 (1200 MHz) & -60.0 & -61 &\\ \hline
\multirow{2}{*}{$180$ GHz without BPF} 
                      & CH1 (600 MHz) & -50.0 & -60 & \multirow{2}{*}{\includegraphics[width=0.1\textwidth]{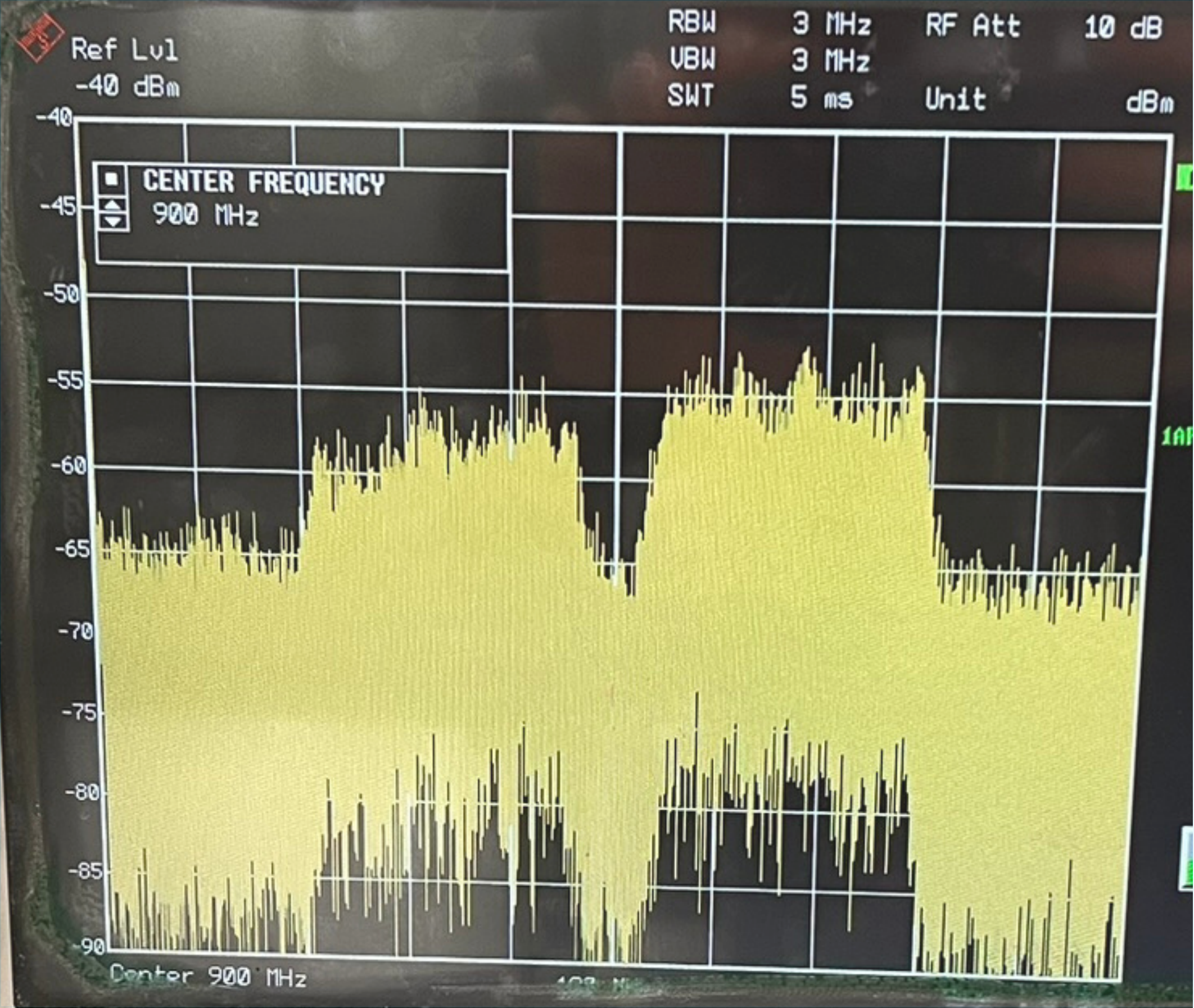}}\\ & CH2 (1200 MHz) & -47.0 & -60 &\\ \hline
\end{tabular}
\end{table*}

A cascaded wideband OFDM transmission is implemented using the RFSoC in a dual-DAC configuration, prior to sub-THz upconversion in a first test. Two independent channels, CH1 and CH2, are centered at 600 MHz and 1.2 GHz, respectively.

Each DAC channel's output power is controlled through the amplifier supply voltage. Measurements show that the output power increases monotonically with supply voltage, transitioning from noise-limited levels (–70 dBm) below 2 V to linear amplification above this threshold. An operating point of 3 V is selected for all subsequent measurements, corresponding to an average output power close to –40 dBm, which ensures linear operation and avoids amplifier saturation. At this bias, the maximum in-band power is measured as –35 dBm for CH1 and –33 dBm for CH2.

Spectral measurements confirm stable wideband operation with clear channel separation. CH1 occupies the frequency range from 375 MHz to 815 MHz, while CH2 spans 975 MHz to 1420 MHz, corresponding to measured spectral envelopes of 350–870 MHz and 930–1460 MHz, respectively. The results demonstrate concurrent wideband transmission across both DAC channels without observable spectral overlap or mutual interference.

Next, the cascaded wideband OFDM implementation is achieved in 180 GHz for the second and third measurements. The IF amplifier voltage is maintained at 3 V, yielding approximately –40 dBm DAC output power, and the FEX-TX and FEX-RX are separated by 13 cm with manual alignment to maximize received power. LO power is 3 dBm for both FEXs, and the RFSoC IF PA is set to 3 V, while the sub-THz receiver LNA operates at 5 V.

\subsection{Spectral Occupancy and Received Power} 

The occupied frequency bands for both channels remain consistent with the previous measurements, but the received power levels are significantly affected by sub-THz upconversion and the presence of BPF. With the 140–220 GHz BPF installed at the FEX-TX front end, CH1 spans 375–815 MHz with in-band power ranging from –66 to –61 dBm, while CH2 spans 975–1420 MHz at approximately –61 dBm. The maximum spectral peak observed within the replicated Nyquist zones is –44 dBm at 1.108 GHz. Compared to the sub-6 GHz baseline (–35 dBm for CH1 and –33 dBm for CH2), the filter introduces attenuation of 28–31 dB, consistent with insertion loss and filtering effects.

Removing the BPF increases the received power across both channels due to the absence of filter attenuation. CH1 now spans 375–815 MHz with in-band power –52 to –49 dBm, while CH2 spans 975–1420 MHz at –49 dBm. The spectral peak rises slightly to –45 dBm at 1.108 GHz. Despite the higher received power, the occupied frequency ranges remain unchanged, demonstrating that the spectral shaping is primarily determined by the OFDM, while the BPF mainly affects amplitude without distorting the frequency coverage.

\subsection{Phase Noise}

Phase noise measurements were performed for both channels at each stage to assess carrier stability. At the sub-6 GHz baseline, CH1 and CH2 exhibited –65 dBc/Hz and –63 dBc/Hz, respectively, at –40 dBm output. Following sub-THz upconversion with the BPF, phase noise slightly increased to –63 dBc/Hz for CH1 and –61 dBc/Hz for CH2, reflecting additional noise contributions from the frequency multipliers and the filter insertion loss. Without the BPF, phase noise improved slightly to –60 dBc/Hz for both channels, consistent with the higher received power and the absence of filter-induced degradation.

The combined results demonstrate that while the BPF reduces overall power and slightly increases phase noise, it does not compromise wideband dual-channel spectral coverage. The DAC-generated OFDM signal maintains its structure, and phase noise remains within acceptable levels for high-frequency sub-THz communication. Table~\ref{tab:power_phasenoise} summarizes the phase noise for all three measurements.

\section{Conclusion}

We demonstrate in this work a real-time OFDM signal generation with $\sim$1.1 GHz of bandwidth with multi–RF-channel, cascaded SDR-based platform achieving an aggregate sampling rate of  2×3.84 GSPS and successfully upconverted signals to 180 GHz. Spectral measurements confirm stable wideband operation with preserved OFDM integrity, while the 140–220 GHz bandpass filter introduces 28–31 dB attenuation and minor phase noise increase without distorting the spectrum. Looking forward, immediate scaling steps involve integrating multiple RFSoCs to increase the number of transmitters, supporting wider aggregate bandwidth.
\bibliographystyle{ieeetr}
\bibliography{refs}
\end{document}